\documentclass[12pt,preprint,preprintnumbers,nofootinbib]{revtex4}
\usepackage{graphicx}
\usepackage{epsfig}    
\usepackage{dcolumn}
\usepackage{bm}
\usepackage{amssymb}
\usepackage{epsfig}    
\usepackage{here}
\newcommand{\scr}[1]{\mbox{\scriptsize{#1}}}
\newcommand{\non}{\nonumber}
\def\gtsim{\mathrel{\hbox{\raise0.2ex
\hbox{$>$}\kern-0.75em\raise-0.9ex\hbox{$\sim$}}}}
\def\ltsim{\mathrel{\hbox{\raise0.2ex
\hbox{$<$}\kern-0.75em\raise-0.9ex\hbox{$\sim$}}}}
\begin{document}
\topmargin -1cm
\begin{flushright}
OU-HET 503\\
KEK-TH-998\\
\end{flushright}
\begin{center}
{\large \bf Electroweak baryogenesis and quantum corrections to the triple Higgs boson coupling}

{{\sc Shinya Kanemura}$^{a,\,}$\footnote{E-mail: kanemu@het.phys.sci.osaka-u.ac.jp}, 
{\sc Yasuhiro Okada}$^{b,c,\,}$\footnote{E-mail: yasuhiro.okada@kek.jp},  
{\sc Eibun Senaha}$^{b,c,\,}$\footnote{E-mail: senaha@post.kek.jp}} \\
{\em $^a$Department of Physics, Osaka University, Toyonaka, Osaka
 560-0043, Japan\\
     $^b$Theory Group, KEK, Tsukuba, Ibaraki 305-0801, Japan\\
     $^c$Department of Particle and Nuclear Physics, 
         the Graduate University for Advanced Studies (Sokendai), Tsukuba, 
         Ibaraki 305-0801, Japan}
\end{center}
\begin{abstract}
A phenomenological consequence of electroweak baryogenesis is studied in connection with
the Higgs physics.
In a two Higgs double model, the first order phase transition can be strong enough to
 allow the electroweak baryogenesis due to the effect of extra Higgs bosons.
We investigate the quantum correction to the triple coupling of the lightest Higgs boson in such a scenario,
and find that the condition of the strong first order phase transition necessarily leads to
the deviation of at least 10\% from the standard model prediction.
Such magnitude of the deviation can be identified in future $e^+e^-$ linear collider experiments.
We also discuss the minimal supersymmetric standard model with the light stop scenario,
and point out that a sizable correction appears in the triple coupling for
 successful electroweak baryogenesis.
\end{abstract}
\maketitle
One of the most challenging problems in particle physics and cosmology is to explain
the baryon asymmetry of the Universe, $n_B/s\sim10^{-10}$~\cite{wmap}, where $n_B$ is
the difference between the number density of baryons and that of anti-baryons, and $s$
is the entropy density. Three conditions
are required for generation of baryon asymmetry starting from a baryon-symmetric
 initial state~\cite{Sakharov:1967dj}: (1)~baryon
number nonconservation, (2)~$C$ and $CP$ violation and (3)~departure from thermal equilibrium. 
These conditions can be in principle satisfied at the electroweak phase transition, because 
baryon number violating processes are efficient at high temperature~\cite{Kuzmin:1985mm}.
In the minimal Standard Model (SM),
however, it turns out to be difficult to generate sufficient baryon asymmetry observed today, 
since the CP violating phase from Cabibbo-Kobayashi-Maskawa matrix is too small and the 
strength of the phase transition is too weak for the current Higgs boson mass bound
($m_h\gtsim114$ GeV~\cite{Barate:2003sz}). 
To realize the baryogenesis at the electroweak phase transition we must extend the minimal SM Higgs sector.
Various scenarios have been considered for electroweak baryogenesis~\cite{ewbg}. 
\\
\indent
In this Letter, we consider the electroweak baryogenesis in the Two Higgs Doublet Model (THDM)~
\cite{ewbg-thdm, simplified-thdm, cline, funakubo}.
In particular, we focus on the connection between cosmology
and collider physics. We study critical temperature 
and the order of phase transition by the use of the one-loop effective potential at finite temperature. 
It is found that the phase transition can be strongly first order enough for baryogenesis to 
take place owing to the non-decoupling effect of the heavy Higgs bosons. We also calculate 
the radiative correction to the triple coupling ($hhh$) of the lightest Higgs boson.  
We find that the deviation of the $hhh$ coupling from the SM value is larger than 
about 10 percent for $m_h=120$~GeV when the necessary condition for the electroweak 
baryogenesis is satisfied. Since the experimental accuracy of the $hhh$ coupling determination 
is expected to be ${\cal O}(10)\%$ from the double Higgs production process
at a future $e^+e^-$ Linear Collider (LC)~\cite{hhh-exp}, 
the self-coupling measurement can provide an important test of the
electroweak baryogenesis scenario
\footnote{
The preliminary result is reported in Ref.~\cite{kos}.
}.
We then extend our consideration to the case of minimal supersymmetric standard model (MSSM).
\\
\indent
In order to create the large enough baryon number asymmetry during the electroweak phase transition, 
we need two conditions on the Higgs potential. First, the phase transition must be strongly first order to 
avoid the washout of the generated baryon number density by the sphaleron process after the phase 
transition~\cite{phi-T}. Secondly, there should be CP violating phases in the Higgs potential at finite temperature, 
which are necessary to generate asymmetry of the fermion and anti-fermion around the bubble wall during 
the phase transition. The CP violating phases from the Higgs sector originate either from explicit or 
spontaneous CP violation. Therefore, connection between the CP phases at the phase transition and 
those in the zero-temperature effective  potential strongly depends on the scenario for the electroweak baryogenesis. 
On the other hand, the strong first-order phase transition is a necessary requirement independent of 
the detail of the mechanism for baryogenesis. Hence, we here focus on the first condition for the finite 
temperature effective potential~\cite{ewpt-thdm}.
\\
\indent
In the THDM the tree-level Higgs potential is given by
\begin{eqnarray}
\non V_{\scr{tree}}&=&m_1^2|\Phi_1|^2+m_2^2|\Phi_2|^2-(m_3^2\Phi_1^\dagger\Phi_2+\mbox{h.c.})\\
&&+\frac{\lambda_1}{2}|\Phi_1|^4+\frac{\lambda_2}{2}|\Phi_2|^4+\lambda_3|\Phi_1|^2|\Phi_2|^2
+\lambda_4|\Phi_1^\dagger\Phi_2|^2+\Bigg[\frac{\lambda_5}{2}(\Phi_1^\dagger\Phi_2)^2
+\mbox{h.c.}\Bigg],\label{higgs-pot}
\end{eqnarray}
where we imposed the softly-broken discrete symmetry as $\Phi_1\rightarrow\Phi_1,~\Phi_2\rightarrow-\Phi_2$ 
to suppress flavor changing neutral current processes~\cite{fcnc} 
\footnote{
According to the types of Yukawa 
interaction we can categorize the THDM into so called Model I and Model II. 
In Model I all fermions receive their masses from only one Higgs field, whereas in Model II the up-type 
quarks receive their masses  from $\Phi_2$ and the down-type quarks and leptons do from $\Phi_1$. 
The following discussion does not depend on the choice of Model I or Model II.}.
Here $m_3^2$ and $\lambda_5$ are generally complex. There is one physical phase which cannot be 
removed by rephasing of the doublet fields, and this becomes the source of explicit CP violation.
Although this phase may be important for generation of baryon asymmetry, 
we neglect it in the following discussion on the electroweak phase transition
\footnote{
This is justified as long as the small phase approximation is valid to solve
the bubble wall equations~\cite{cline, funakubo}.}.
In the limit of CP invariance there are five physical states; 
CP-even ($h, H$), CP-odd ($A$) and charged Higgs bosons ($H^\pm$). Here, $h$ is lighter than $H$. 
We introduce mixing angles $\alpha$ and $\beta$. The angle $\alpha$ is defined as the mixing angle 
between CP-even Higgs bosons, and $\tan\beta~(=\langle\Phi_2^0\rangle/\langle\Phi_1^0\rangle)$ 
is the ratio of the vacuum expectation values (VEVs) of the two Higgs fields. 
These VEVs satisfy $v=\sqrt{2}\sqrt{\langle\Phi_1^0\rangle^2+\langle\Phi_2^0\rangle^2}\simeq246$~GeV. 
In addition, we introduce $M^2~(=m_3^2/\sin\beta\cos\beta)$, which characterizes the soft breaking 
scale of the discrete symmetry. There are two cases concerning the decoupling property of the 
additional heavy Higgs bosons. When $M^2\gg v^2$, all the masses of the heavy Higgs bosons are 
approximately given by $M$, and the properties of the lightest Higgs boson become similar to those of
 the SM Higgs boson. In this case the loop effect due to the heavy Higgs bosons is suppressed by the factor of $1/M^2$. 
On the contrary, in the opposite limit where all the heavy Higgs bosons receive their masses from the VEVs, 
 the loop effect due to the heavy Higgs bosons does not necessarily decouple in the large mass limit.  
\\
\indent
In the following, we assume $m_1=m_2$ and
$\lambda_1=\lambda_2$ in order to simplify the analysis of the phase transition. In this case, the field direction 
relevant to the electroweak phase transition is reduced to 
\begin{equation}
\langle\Phi_1\rangle=\langle\Phi_2\rangle=
\left(
\begin{array}{c}
0\\
\frac{1}{2}\varphi
\end{array}
\right).
\end{equation}
This case corresponds to $\sin(\alpha-\beta)=-1$ and $\tan\beta=1$. This parameter choice has been
proposed for a successful scenario of electroweak baryogenesis~\cite{simplified-thdm, cline}
 to avoid the complexity with two stage phase transition~\cite{two-stage}.\\
\indent
The quantum correction of the extra Higgs bosons to the $hhh$ coupling is calculated in the THDM
 using both effective potential and diagrammatic methods in Refs. \cite{kosy, kkosy}.
The effective potential is given by
$V_{\scr{eff}}[\varphi] = V_{\scr{tree}}[\varphi] + \Delta V[\varphi]$, where $\Delta V[\varphi] $ is expressed as
\begin{eqnarray}
   \Delta V[\varphi] = \frac{1}{64\pi^2} \sum_f 
                     N_{c_f} N_{s_f} (-1)^{2 s_f}  
                     (M_f[\varphi])^4 
             \left\{ \ln \frac{(M_f[\varphi])^2}{Q^2}
                         -\frac{3}{2}\right\}. 
\label{one-eff}
\end{eqnarray}
Here $N_{c_f}$ is the color number,   
$s_f$ ($N_{s_f}$) is the spin (degree of freedom) of the field $f$  
in the loop, $M_f[\varphi]$ is the field dependent mass of $f$, 
and $Q$ is a renormalization scale. 
 In Ref.~\cite{kkosy}, 
 we found that the deviation of the $hhh$ coupling constant from SM value can be ${\cal O}(100)\%$ due to the 
 non-decoupling effect of the heavy Higgs bosons. When the Higgs couplings to gauge bosons are SM-like at the tree-level 
 ($\sin(\alpha-\beta)\simeq-1$), the leading loop contributions of the heavy Higgs bosons and the top quark are 
 extracted as
\begin{eqnarray}
\non\lambda_{hhh}^{\scr{eff}}(\mbox{THDM})&\simeq&\frac{3m_h^2}{v}\Bigg[1+\frac{m_H^4}{12\pi^2m_h^2v^2}\Bigg(1-\frac{M^2}{m_H^2}\Bigg)^3+\frac{m_A^4}{12\pi^2m_h^2v^2}\Bigg(1-\frac{M^2}{m_A^2}\Bigg)^3\\
&&\hspace{5cm}+\frac{m_{H^\pm}^4}{6\pi^2m_h^2v^2}\Bigg(1-\frac{M^2}{m_{H^\pm}^2}\Bigg)^3-\frac{m_t^4}{\pi^2m_h^2v^2}\Bigg].\label{lambdahhh}
\end{eqnarray}
It is easily seen that the effects of the heavy Higgs boson loops are enhanced 
by $m_\Phi^4$ ($\Phi=H, A, H^\pm$) when $M^2$ is zero. These effects do not decouple in the 
large mass limit $m_\Phi \rightarrow \infty$ and yields the large deviation of $hhh$ coupling from the SM prediction. \\
\indent
In order to study the electroweak phase transition, we consider the effective potential at finite temperature: 
\begin{eqnarray}
V_{\scr{eff}}(\varphi, T)=V_{\scr{tree}}(\varphi)+\Delta V(\varphi)+\Delta V^{(T)}(\varphi, T),\label{eff_T}
\end{eqnarray}
where the finite temperature contribution is expressed as
\begin{eqnarray}
\Delta V^{(T)}(\varphi, T)&=&\frac{T^4}{2\pi^2}\Big[\sum_{i=\scr{bosons}}n_iI_B(a^2_i)+n_tI_F(a^2_t)\Big],\label{eff_T}
\end{eqnarray}
with
\begin{equation}
I_{B,F}(a^2)=\int_0^\infty dx~x^2\log(1\mp e^{-\sqrt{x^2+a^2}}).
\end{equation}
In the above expressions, $n_i$ is the degree of freedom of the particle $i$: $n_W=6,~n_Z=3,~n_t
=-12,~n_h=n_H=n_A=1,~n_{H^\pm}=2$, $m_i(\varphi)$ is the field dependent mass and $a_i(\varphi)=m_i(\varphi)/T$. \\
\indent
Before going to the numerical evaluation, we discuss qualitative features of the phase transition by
using the high temperature expansion.
For high temperature $T\gg m_i $, we may expand the integrals $I_{B,F}^{}$ analytically~\cite{dj}.
When $m_\Phi^2\gg m_h^2, M^2$ ($\Phi=H, A, H^\pm$), the field dependent masses of the
heavy Higgs bosons can be written as $m_\Phi^2(\varphi)\simeq m_\Phi^2\varphi^2/v^2$.
In the high temperature approximation, the effective potential at finite temperature of Eq.~(\ref{eff_T})
is expanded as~\cite{analy-exp}
\begin{equation}
V_{\scr{eff}}(\varphi, T)\simeq D(T^2-T^2_0)\varphi^2-ET|\varphi|^3+\frac{\lambda_{T}}{4}\varphi^4+\cdots,
\label{hte-pot}
\end{equation}
where
\begin{eqnarray}
D &=& \frac{1}{24 v^2} (6 m_W^2 + 3 m_Z^2 + 6 m_t^2 
  + m_H^2 + m_A^2 + 2 m_{H^\pm}^2),\\
T_0^2 &=&  \left\{ \frac{1}{4} m_h^2 - 
 \frac{1}{32\pi^2v^2} 
 (6 m_W^4 + 3 m_Z^4 - 12 m_t^4 + m_H^4 + m_A^4 + 2 m_{H^\pm}^4)
\right\}/D,\\
E &=& \frac{1}{12 \pi v^3} (6 m_W^3 + 3 m_Z^3  
 + m_H^3 + m_A^3 + 2 m_{H^\pm}^3),\label{trilinear}\\
\lambda_T &=& 
 \frac{m_h^2}{2 v^2} 
 \left[ 1 
- \frac{1}{8\pi^2 v^2 m_h^2} 
 \left\{ 
 6 m_W^4 \log \frac{m_W^2}{\alpha_B^{} T^2}  
+  3 m_Z^4 \log \frac{m_Z^2}{\alpha_B^{} T^2}  
- 12 m_t^4 \log \frac{m_t^2}{\alpha_F^{} T^2}  \right.\right.\nonumber\\
&&\qquad \left.\left. 
+  m_H^4 \log \frac{m_H^2}{\alpha_B^{} T^2}  
+  m_A^4 \log \frac{m_A^2}{\alpha_B^{} T^2}  
+ 2  m_{H^\pm}^4 \log \frac{m_{H^\pm}^2}{\alpha_B^{} T^2}  
\right\}   
\right],
\end{eqnarray}
where $\log\alpha_B^{}=2 \log 4 \pi - 2 \gamma_E^{}$, $\log\alpha_F^{}=2 \log \pi - 2 \gamma_E^{}$
and $\gamma_E^{}$ is the Euler constant.
The first order phase transition becomes possible due to appearance of the cubic term which originates from the boson loops
at finite temperature.  
Unlike the SM, the coefficient of the cubic term $E$ can receive a large contribution from the extra Higgs bosons.
At the critical temperature $T_c$,  the potential has two degenerate minima at $\varphi=0$
and $\varphi_c=2ET_c/\lambda_{T_c}$, where $\lambda_{T_c}$ is the quartic coupling constant at $T_c$. 
In order not to wash out the created baryon number density after the electroweak phase transition, 
we have to require that the sphaleron process should be sufficiently suppressed.
The most reliable condition has been obtained from the lattice simulation study~\cite{sph}.
It is expressed as
\begin{equation}
\frac{\varphi_c}{T_c}=\frac{2E}{\lambda_{T_c}}\gtsim1.\label{sph}
\end{equation}
For $m_h=120$~GeV, this condition can be satisfied when the masses of the heavy Higgs bosons are
above 200 GeV. We can see from Eq.~(\ref{lambdahhh}) that the correction to the $hhh$ coupling can
be large in such a parameter region. Although the high temperature expansion gives a qualitative description
of the phase transition, the approximation breaks down when the masses of the heavy Higgs bosons
become larger than the critical temperature. We therefore evaluate the effective potential numerically
and search the parameter space where the condition (\ref{sph}) is satisfied.\\
\indent
\begin{figure}
\epsfxsize=10cm
\centerline{\epsfbox{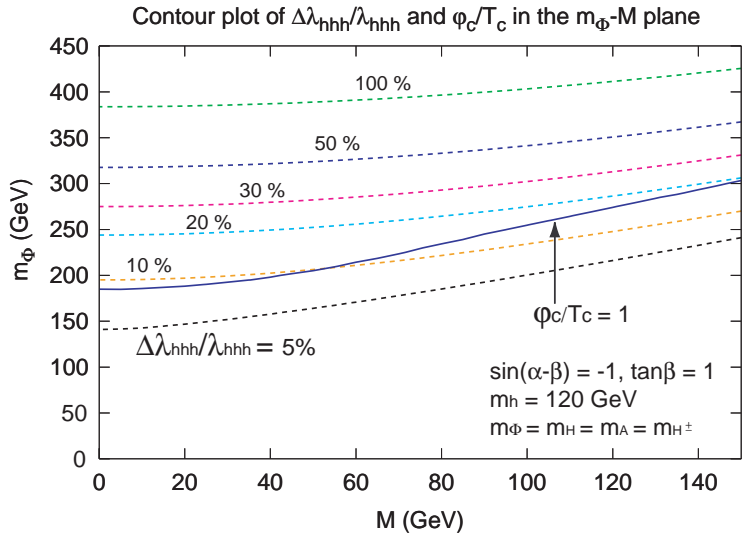}}
\caption{The straight line stands for the critical line which satisfied the condition,
$\varphi_c/T_c=1$. The dashed lines are the deviation of hhh coupling from the SM value,
where  $\Delta \lambda_{hhh}^{\scr{THDM}}\equiv \lambda_{hhh}^{\scr{eff}}
(\mbox{THDM})-\lambda_{hhh}^{\scr{eff}}(\mbox{SM})$.}  
\label{figA}
\end{figure}
In Fig.~\ref{figA}, 
we show the parameter region where the necessary condition of the electroweak
baryogenesis in Eq.~(\ref{sph}) is satisfied in the $m_\Phi^{}$-$M$ plane.
We take 
$\sin(\alpha-\beta)=-1$, $\tan\beta=1$ and $m_h=120$~GeV. For the heavy Higgs boson mass, we assume
$m_H^{}=m_{A}^{}=m_{H^\pm}^{} (\equiv m_\Phi^{})$ to avoid the constraint on the $\rho$ parameter from the
LEP precision data~\cite{pdg}.
In the numerical evaluation, we take into account the ring summation for
the contribution of the Higgs bosons to the effective potential at finite temperature~\cite{dj,daisy}. 
 For fixed values of $m_\Phi^{}$ and $M$, we calculate the effective 
potential (\ref{eff_T}) varying the temperature $T$ and determine the critical temperature $T_c^{}$ of the
first-order phase transition and the expectation value $\varphi_c^{}$ at $T_c^{}$. 
We can see that the phase transition becomes strong enough for the baryogenesis when the masses 
of the heavy Higgs bosons are larger than about 200 GeV. For larger values of $M$, greater
$m_\Phi^{}$ are required to satisfy the condition (\ref{sph}). In this figure we also plot the contour of the
magnitude of the deviation in the $hhh$ coupling from the SM value.
We define the deviation $\Delta \lambda_{hhh}^{\scr{THDM}}/
\lambda_{hhh}^{\scr{eff}}(\mbox{SM})$ by $\Delta \lambda_{hhh}^{\scr{THDM}}\equiv
\lambda_{hhh}^{\scr{eff}}(\mbox{THDM})-\lambda_{hhh}^{\scr{eff}}(\mbox{SM})$.
We calculated the deviation at the one loop level in the on-shell scheme which gives a better approximation
 than the use of Eq. (\ref{lambdahhh})~\cite{kkosy, kosy}.
We can easily see
that the magnitude of the deviation is significant ($\gtsim10\%$) in the parameter region where the electroweak
 baryogenesis is possible. Such magnitude of the deviation can be detected at a future LC experiment. \\
\indent
Next we discuss a scenario of electroweak baryogenesis in the MSSM.
In this case, the strong first order phase transition can be induced by the loop effect
of the light stop in the finite temperature effective potential~\cite{cqw96, ewbg-mssm}.
We examine the loop effect of the light stop on the $hhh$ coupling in this scenario.
In the following, we only consider the finite and zero temperature effective potentials in a simple approximation
to understand the qualitative feature.
\\
\indent
In order to realize the successful baryogenesis under the constraint from the $\rho$ parameter
and the lower bound of the lightest Higgs boson mass, we have to take the 
$M_Q^2 \gg M_U^2, m_t^2$, where $M_Q$ and $M_U$ are soft breaking masses associated with
left and right stops, respectively. The eigenvalues of the field dependent mass matrix are expressed as  
\begin{eqnarray}
m_{\tilde{t}_1}^2(\varphi,\beta) 
&=& M_U^2 + D_R^2 
+ \frac{h_t^2 \sin^2\beta}{2} 
\left( 1 - \frac{|X_t|^2}{M_Q^2} \right) \varphi^2,\\
m_{\tilde{t}_2}^2(\varphi,\beta) 
&=& M_Q^2 + D_L^2 
+ \frac{h_t^2 \sin^2\beta}{2} 
\left( 1 + \frac{|X_t|^2}{M_Q^2} \right) \varphi^2
 \simeq M_Q^2, 
\end{eqnarray}
where $X_t=A_t + \mu \cot\beta$ with $A_t$ and $\mu$ being the stop trilinear coupling
and the higgsino mass term, respectively, and $D_L$ and $D_R$ are corresponding D-term
contributions. 
In the high temperature expansion,  the contribution to the coefficient of the cubic term in the 
finite temperature effective potential (\ref{hte-pot}) is given by \cite{cqw96}
\begin{eqnarray}
\Delta E_{\tilde{t}_1} 
 \simeq + \frac{1}{6\pi} \frac{N_c m_t^3}{v^3}
\left( 1 - \frac{|X_t|^2}{M_Q^2} \right)^{3/2},
\end{eqnarray}
for $M_U^2+D_R^2\sim0$.
It is known that electroweak baryogenesis is possible due to the stop contribution
if the lightest Higgs mass $m_h$ is close to the current experimental lower bound.
We also evaluate the one loop effect of the light stop to the $hhh$ coupling, and 
extract the leading contribution as
\begin{eqnarray}
\frac{\Delta \lambda_{hhh}(MSSM)}{\lambda_{hhh}(SM)} 
 \simeq \frac{N_c m_t^4}{6 \pi^2 v^2 m_h^2}
\left( 1 - \frac{|X_t|^2}{M_Q^2} \right)^3, 
\end{eqnarray}
where $m_h$ is the one-loop corrected mass of the lightest Higgs boson.
Combining the above equations, we can derive
\begin{equation}
\frac{\Delta \lambda_{hhh}(MSSM)}{\lambda_{hhh}(SM)} 
 \simeq \frac{6v^4}{m_t^2m_h^2N_c}(\Delta E_{\tilde{t}_1})^2.
\end{equation}
From the condition (\ref{sph}), the deviation in the $hhh$ coupling from the SM value is estimated to be
larger than $\sim6\%$. Although the above calculation needs to be improved, 
this example suggests that a large correction in the $hhh$ coupling
is a general feature of successful electroweak baryogenesis.
\\
\indent
We have studied the electroweak phase transition by using the effective potential at finite temperature
 in the THDM and the MSSM. 
 In the THDM, it is found that the phase transition can be strongly first order enough for successful
electroweak baryogenesis due to the non-decoupling effects of the heavy Higgs boson loops at
 finite temperature. Such non-decoupling effects also affect the triple Higgs self-coupling constant of the
lightest Higgs boson at the one-loop level. We found that if the baryogenesis occurs at the electroweak 
phase transition, the $hhh$ coupling constant deviates from the SM prediction at least $\sim$10\% in the THDM.
The measurement of the Higgs self-coupling therefore can provide an interesting interplay
between cosmology and collider physics.
\\[1cm]
%
\textbf{ACKNOWLEDGEMENTS}\\
We would like to thank Koichi Funakubo for useful discussions.
The work of YO was supported in part by a Grant-in-Aid of the Ministry 
of Education, Culture, Sports, Science, and Technology, Government of
Japan, Nos.~13640309, 13135225, and 16081211.

\end{document}